\documentclass{ckm}                 

\def\bra#1{\left\langle #1\right|}
\def\ket#1{\left| #1\right\rangle}
\newcommand{\bers}{\begin{eqnarray*}}
\newcommand{\eers}{\end{eqnarray*}}
\newcommand{\bt}{\begin{itemize}}
\newcommand{\et}{\end{itemize}}
\def\beq{\begin{equation}}
\def\eeq{\end{equation}}
\def\bea{\begin{eqnarray}}
\def\eea{\end{eqnarray}}
\def\nn{\nonumber}
\def\sla#1{\raise.15ex\hbox{$/$}\kern-.57em #1}

\def\sss{\scriptscriptstyle}

\def\bd{B_d^0}

\def\barp{{\raise.35ex\hbox
{${\sss (}$}}---{\raise.35ex\hbox{${\sss )}$}}}
\def\bdbarp{\hbox{$B_d$\kern-1.4em\raise1.4ex\hbox{\barp}}}
\def\bsbarp{\hbox{$B_s$\kern-1.4em\raise1.4ex\hbox{\barp}}}
\def\ks{K_{\sss S}}

\def\roughly#1{\mathrel{\raise.3ex\hbox
{$#1$\kern-.75em\lower1ex\hbox{$\sim$}}}}

\def\npb#1#2#3{{ Nucl.\ Phys.} {\bf B#1}, #3 (#2)}
\def\plb#1#2#3{{ Phys.\ Lett.} {\bf #1B}, #3 (#2)}
\def\prd#1#2#3{{ Phys.\ Rev.} {\bf D#1}, #3 (#2)}
\def\newprd#1#2#3{{ Phys.\ Rev.} {\bf D#1}, #3 (#2)}
\def\prl#1#2#3{{ Phys.\ Rev.\ Lett.} {\bf #1}, #3 (#2)}

\def\zpc#1#2#3{{ Zeit.\ Phys.} {\bf C#1}, #3 (#2)}

\confname{Workshop on the CKM Unitarity Triangle, IPPP Durham, April
  2003 \quad \quad \quad \quad \quad \quad \quad UdeM-GPP-TH-03-111 }

\title{T violation in $B \to V V$ Decays in the SM and Beyond}

\author{Alakabha Datta\addressmark{a} and David London
\addressmark{b}
}

\address[a]{
{\it Department of Physics, University of Toronto,}\\
{\it 60 St.\ George Street, Toronto, ON, Canada M5S 1A7}\\
}
\address[b]{
{\it Laboratoire Ren\'e J.-A. L\'evesque, 
Universit\'e de Montr\'eal,}\\
{\it C.P. 6128, succ. centre-ville, Montr\'eal, QC,
Canada H3C 3J7}
}

\begin{document}

\begin{abstract}
In this talk, we describe T-violating triple-product correlations
(TP's) in $B \to V_1 V_2$ decays. We point out that TP's in the SM are
generally tiny. It is only in a few decays with excited vector mesons
in the final state that measurable TP's may be obtained. On the other
hand, TP's in models beyond the SM can be large, and hence TP
correlations are excellent probes of new physics.
\end{abstract}

\maketitle


\section{Triple Products}

The study of CP violation in the $B$ system is very useful for
understanding the flavour sector of the standard model (SM) and for
searching for new physics. Most CP-violation studies involve
mixing-induced CP-violating asymmetries in neutral $B$ decays and
direct CP asymmetries \cite{CPreview}. In any local and
Lorentz-invariant field theory, CP violation implies T violation from
CPT conservation. The study of T violation can therefore yield further
information on CP- and T-violating phases in the SM or in models of
new physics.

In general T violation is studied via triple-product
correlations. These triple products (TP's) take the form $\vec v_1
\cdot (\vec v_2 \times \vec v_3)$, where each $v_i$ is a spin or
momentum, and are odd under time reversal. One can define a TP
asymmetry as
\beq
A_{\sss T} \equiv 
{{\Gamma (\vec v_1 \cdot (\vec v_2 \times \vec v_3)>0) - 
\Gamma (\vec v_1 \cdot (\vec v_2 \times \vec v_3)<0)} \over 
{\Gamma (\vec v_1 \cdot (\vec v_2 \times \vec v_3)>0) + 
\Gamma (\vec v_1 \cdot (\vec v_2 \times \vec v_3)<0)}} ~,
\label{Toddasym}
\eeq
where $\Gamma$ is the decay rate for the process in question.

Unfortunately, strong phases can produce a nonzero value of $A_{\sss
T}$, even if the weak phases are zero (i.e.\ there is no CP
violation).  Hence, to search for a true T-violating signal one should
compare $A_{\sss T}$ with ${\bar A}_{\sss T}$, where ${\bar A}_{\sss
T}$ is the T-odd asymmetry measured in the CP-conjugate decay process
\cite{Valencia}.

Like CP asymmetries, TP asymmetries are nonzero only if there are two
interfering decay amplitudes. However, there is an important
difference between the two. Denoting $\phi$ and $\delta$ as the
relative weak and strong phases, respectively, between the two
interfering amplitudes, the signal for direct CP violation can be
written
\beq
{\cal A}_{\sss CP}^{dir} \propto \sin\phi \sin\delta ~,
\label{Adirform}
\eeq
while that for the (true T-violating) TP asymmetry is given by
\beq
{\cal A}_{\sss T}  \propto \sin\phi \cos\delta ~.
\label{TPform}
\eeq
Since strong phases in $B$ decays are expected to be small due to the
heavy $b$-quark mass, it is likely that triple-product asymmetries
will be easier to measure than direct CP asymmetries.

In the rest frame of the $B$, the TP for the process $B \to V_1 V_2$
takes the form ${\vec q} \cdot ({\vec\varepsilon}_1 \times
{\vec\varepsilon}_2)$, where ${\vec q}$ is the momentum of one of the
final vector mesons, and ${\vec\varepsilon}_1$ and
${\vec\varepsilon}_2$ are the polarizations of $V_1$ and $V_2$. TP
signals in the $B$ system were studied several years ago by Valencia
\cite{Valencia}, and several general studies of $B\to V_1 V_2$ decays
within the SM were subsequently performed \cite{KP,DDLR,CW,Chiang}. In
Ref.~\cite{DL}, upon which this talk is based, these past analyses are
updated and extended, and the effects of physics beyond the SM on TP
asymmetries are considered.

We write the decay amplitude for $B(p) \to V_1(k_1,\varepsilon_1) +
V_2(k_2,\varepsilon_2)$ as follows:
\bea
M & = & a \, \varepsilon_1^* \cdot \varepsilon_2^* + {b \over m_B^2}
(p\cdot \varepsilon_1^*) (p\cdot \varepsilon_2^*) \nn\\
& & \qquad\qquad\qquad + \, i {c \over m_B^2}
\epsilon_{\mu\nu\rho\sigma} p^\mu q^\nu \varepsilon_1^{*\rho}
\varepsilon_2^{*\sigma} ~,
\label{abcdefs}
\eea
where $q\equiv k_1 - k_2$. (Note that we have normalized terms with a
factor $m_B^2$, rather than $m_1 m_2$ as in Ref.~\cite{Valencia}.
With the above normalization, each of $a$, $b$ and $c$ is expected to
be the same order of magnitude.)  The quantities $a$, $b$ and $c$ are
complex and will in general contain both CP-conserving strong phases
and CP-violating weak phases. In $|M|^2$, a triple-product
correlation arises from interference terms involving the $c$
amplitude, and will be present if ${\rm Im}(a c^*)$ or ${\rm Im}(b
c^*)$ is nonzero. As discussed above, true T-violating effects are
obtained by comparing the triple product measured in $B\to V_1 V_2$
with that obtained in the CP-conjugate process. In order to
experimentally measure TP's in $B\to V_1 V_2$ decays, it is necessary
to perform an angular analysis to extract the terms ${\rm Im}(a c^*)$
and ${\rm Im}(b c^*)$.

\section{T violation in the Standard Model}

To study non-leptonic $B$ decays one starts with the SM effective
hamiltonian for $B$ decays \cite{BuraseffH}:
\bea
H_{eff}^q &=& {G_F \over \protect \sqrt{2}}
[V_{fb}V^*_{fq}(c_1O_{1f}^q + c_2 O_{2f}^q) \nn\\
&& - \sum_{i=3}^{10}(V_{ub}V^*_{uq} c_i^u
+V_{cb}V^*_{cq} c_i^c +V_{tb}V^*_{tq} c_i^t) O_i^q]\nn\\
&& + \, h.c.,
\label{Heff}
\eea
where the superscript $u$, $c$, $t$ indicates the internal quark, $f$
can be the $u$ or $c$ quark, and $q$ can be either a $d$ or $s$ quark.
Using the notation adopted in Ref.~\cite{datta-he}, we write the
operators $O_i^q$ as
\bea
O_{f1}^q &=& \bar q_\alpha \gamma_\mu Lf_\beta\bar
f_\beta\gamma^\mu Lb_\alpha \nonumber\\
O_{2f}^q  &=& \bar q
\gamma_\mu L f\bar
f\gamma^\mu L b\,\nn\\
O_{3,5}^q &=&\bar q \gamma_\mu L b
\bar q' \gamma^\mu L(R) q'\nonumber\\
O_{4,6}^q  &=& \bar q_\alpha
\gamma_\mu Lb_\beta
\bar q'_\beta \gamma^\mu L(R) q'_\alpha \nonumber\\
O_{7,9}^q &=& {3\over 2}\bar q \gamma_\mu L b  e_{q'}\bar q'
\gamma^\mu R(L)q'\nonumber\\
O_{8,10}^q &=& {3\over 2}\bar q_\alpha
\gamma_\mu L b_\beta
e_{q'}\bar q'_\beta \gamma^\mu R(L) q'_\alpha, 
\eea
where $R(L) = 1 \pm \gamma_5$, and $q'$ is summed over $u$, $d$, $s$,
$c$. $O_2$ and $O_1$ are the tree-level and QCD-corrected operators,
respectively. $O_{3-6}$ are the strong gluon-induced penguin
operators, and operators $O_{7-10}$ are due to $\gamma$ and $Z$
exchange (electroweak penguins), and box diagrams at loop level. The
important point here is that all SM operators involve a left-handed
$b$-quark.

Using factorization to calculate the amplitude for $B\to V_1 V_2$, we
arrive at some general conclusions for TP's within the SM \cite{DL}.
The first observation follows from the fact that TP's are {\it
kinematical} CP-violating effects \cite{Kayser}. In order to observe
TP asymmetries, it is not enough just to have two decay amplitudes
with a relative weak phase. What one really needs is two different
{\it kinematical} amplitudes with a relative weak phase.Thus, even
though there are two amplitudes with different strong and weak phases,
decays like $ B \to D^* {\bar D}^*$ will not produce a TP asymmetry
because the two amplitudes are not kinematically different. In fact,
many SM processes have only one kinematically distinct amplitude
within factorization, and so all TP asymmetries vanish for these
decays.

The second point is that, even with two distinct kinematic amplitudes,
TP asymmetries in the standard model are generally suppressed by
flavour symmetries and by the fact that TP's involve transverse
polarization amplitudes that are mass-suppressed relative to the
longitudinal polarization amplitude \cite{DL}. (This is in contrast to
$\Lambda_b$ decays where measurable TP asymmetries may be obtained in
the SM \cite{SMlambdab}.) One can avoid TP suppressions in $B$ decays
by considering excited vector mesons in the final state. Decays to
radially-excited states were studied in Ref.~\cite{dattalip2}, and for
some of these decays it is possible to have observable TP asymmetries
in the SM \cite{DL}. In Ref.~\cite{DL}, the list of decays which
exhibit TP asymmetries in the SM is presented. This list includes
several modes that were not discussed in earlier work. The bottom line
is that T violation in SM is tiny in most decays, and therefore any
observation of a large T-violation signal would be a clear indication
of new physics.

\section{T violation with New Physics}

As discussed in the previous section, most triple-product asymmetries
in $B\to V_1 V_2$ decays in the SM are predicted to be very small.
The important question is then: can such TP asymmetries be large with
new physics? Also, what kind of new physics can lead to large TP
asymmetries?

It is fairly straightforward to see how new physics can produce a
large T-violating asymmetry where the SM predicts little or no T
violation. The essential point is that the effective SM Hamiltonian
involves only a left-handed $b$-quark, and so contains only $(V-A)
\times (V-A)$ and $(V-A) \times (V+A)$ operators. In other words it is
very difficult to generate two kinematically different amplitudes in
the SM. On the other hand, many models of new physics can couple to
the right-handed $b$-quark, producing $(V+A) \times (V-A)$ and/or
$(V+A) \times (V+A)$ operators. These new-physics operators will
produce different kinematical amplitudes, leading to different phases
for $a$, $b$ and $c$ [Eq.~(\ref{abcdefs})], thus giving rise to a TP
asymmetry. Hence a large measured TP in $B$ decays will not only be a
smoking-gun signal for new physics but will also signal the presence
of nonstandard operators, specifically those involving a right-handed
$b$-quark. In fact, as was shown in Ref.~\cite{NPlambdab}, by studying
TP's in several modes, one can test various models of new physics.
 
To demonstrate the effect of new physics in $B$ decays, we concentrate
on the decay $B \to \phi K^*$. The SM predicts that the indirect CP
asymmetries in $\bd(t) \to J/\psi \ks$ and $\bd(t) \to \phi\ks$ are
expected to be equal, both measuring $\sin 2\beta$. Any difference
between these two measurements should be at most at the level of
0($\lambda^2$), where $\lambda \sim 0.2$. However, at present this
does not appear to be the case. The world averages for these
measurements are \cite{Jpsiks,phiks}:
\bea
\sin 2 \beta ~~[J/\psi \ks] & = & 0.734 \pm 0.054 ~, \nn\\
\sin 2 \beta ~~[\phi \ks] & = & -0.39 \pm 0.41 ~.
\label{phiKsresult}
\eea
Decays that have significant penguin contributions are naturally
likely to be affected by physics beyond the SM \cite{Gross}. As
pointed out in Ref.~\cite{LonSoni}, $\bd \to \phi \ks$ is sensitive to
new physics because it is a pure $b\to s$ penguin decay. The recent CP
measurements in $ B \to \phi K_s$ have led to several recent attempts
to understand the data with new-physics scenarios
\cite{phiKsNP,dattarparity,bhaskar}. (One can also look for
new-physics effects through the measurement of $\sin 2\beta$ in the
decay $B \to \eta^{\prime} K$. However, these decays have large
branching ratios and other complications \cite{dattalip3}, making the
search for new physics in these modes a bit problematic.)

As an example, here we focus on one particular new-physics model, that
of supersymmetry with R-parity violation \cite{dattarparity}. However,
we emphasize that the analysis can be easily extended to other models
of new physics. Assuming that R-parity-violating SUSY is the
explanation for the CP measurements in $\bd(t) \to \phi \ks$, we
estimate here the expected TP asymmetries in $B \to \phi K^*$
\cite{LSS2}.

For the $b \to s {\bar{s}}s$ transition, the relevant terms in the
R-parity-violating SUSY Lagrangian are
\bea
L_{eff} & =& \frac{\lambda^{\prime}_{i32} \lambda^{\prime*}_{i22}} { 4
m_{ \widetilde{\nu}_i}^2} \bar s (1+\gamma_5) s \, \bar {s}
(1-\gamma_5) b\nonumber\\
&& + \, \frac{\lambda^{\prime}_{i22} \lambda^{\prime*}_{i23}}
{ 4 m_{ \widetilde{\nu}_i}^2} \bar s (1-\gamma_5) s \, \bar {s}
(1+\gamma_5) b ~.
\eea
(We refer to Ref.~\cite{dattarparity} for a full explanation of SUSY
with R-parity violation.) The amplitude for $B \to \phi K^{*}$,
including the new-physics contributions, can then be written as
\cite{DL}
\beq
A[B \to \phi K^*] = \frac{G_F}{\sqrt{2}} [(X+X_1) P_{\phi}+X_2
Q_{\phi}] ~,
\eeq
with
\bea
X & = & - \sum_{q=u,c,t}V_{qb}V_{qs}^* \left[ (a_3^q+a_4^q+a_5^q) \right. \nn\\
& & \left. -\frac{1}{2}(a_7^q+a_9^q+a_{10}^q) \right] ~, \nn\\
X_1 &= & -\frac{\sqrt{2}}{G_F} \frac{\lambda^{\prime}_{i32}
\lambda^{\prime*}_{i22}} {24 \, m_{ \widetilde{\nu}_i}^2} ~, \nn\\
X_2 & = & -\frac{\sqrt{2}}{G_F} \frac{\lambda^{\prime}_{i22}
  \lambda^{\prime*}_{i23}} {24 \, m_{ \widetilde{\nu}_i}^2} ~, \nn\\
P_{\phi} & = & m_{\phi}g_{\phi}\varepsilon^{*\mu}_{\phi} \bra{K^*}
 \bar{s} \gamma_{\mu}(1-\gamma_5)b \ket{B} ~, \nn\\
Q_{\phi} & = & m_{\phi}g_{\phi}\varepsilon^{*\mu}_{\phi} \bra{K^*}
 \bar{s} \gamma_{\mu}(1+\gamma_5)b \ket{B} ~.
\label{Bphi-K}
\eea
For $\bd \to \phi \ks$ it is the combination $X_1+X_2$ which
contributes \cite{dattarparity}, and we can define the quantity $X_R$
via
\beq
X_1+X_2=\frac{\sqrt{2}}{G_F}\frac{X_R}{12M^2}e^{i\phi} ~,
\eeq
where $\phi$ is the weak phase in the R-parity-violating couplings,
and $M$ is a mass scale with $M \sim m_{\widetilde{\nu}_i}$. In order
to reproduce the CP-violating $\bd(t) \to \phi \ks$ measurement in
Eq.~(\ref{phiKsresult}), one requires $|X_R| \sim 1.5 \times 10^{-3}$
for $M=100$ GeV, along with a value for the phase $\phi$ near ${\pi
\over 2}$. In our calculations of TP's in $B \to \phi K^{*}$ we make
the simplifying assumption that $X_1=X_2$, and choose $\phi={\pi \over
2}$.

We present our results in Table.~\ref{Bdphi-K}. These results hold for
both neutral and charged $B$ decays. The predicted branching ratio for
$B \to \phi K^*$ is slightly larger than the measured branching ratios
$BR(B^+ \to \phi K^{*+}) = 10^{+5}_{-4} \times 10^{-6}$ and $BR(\bd
\to \phi K^{*0}) = 9.5^{+2.4}_{-2.0} \times 10^{-6}$ \cite{PDG}, but
it is well within the theoretical uncertainties of the
calculation. The important result is that we expect {\it very large}
(15--20\%) TP asymmetries for these decays, as well as for those with
radially-excited final states.

As emphasized above, these large TP asymmetries are not unique to
supersymmetry with R-parity violation. One expects to find large TP
asymmetries in many other models of physics beyond the SM. The
measurement of such TP asymmetries would not only reveal the presence
of new physics, but more specifically it would point to new physics
which includes large couplings to the right-handed $b$-quark.

\begin{table}[thb] 
\begin{center} 
\begin{tabular}{|c|c|c|} 
\hline 
Process &  BR & $A_T^{(1)}$ \% \\ 
\hline 
$B \to \phi K^{*}$ & $16.7~(17.4) \times 10^{-6}$ &$-$16.3 ($-$15.6) \\
\hline 
$B \to \phi^{\prime} K^{*}$ & $19.1~(20.7) \times 10^{-6}$ &$-$21.0
 ($-$19.3) \\
\hline 
$B \to \phi K^{* \prime}$ & $28.0~(28.9) \times 10^{-6}$ &$-$15.4
 ($-$14.8) \\
\hline 
\end{tabular}
\end{center}
\caption{Branching ratios (BR) and triple-product asymmetries
($A_{T}^{(1)}$) for $B \to \phi K^{*}$ and excited states, for $N_c=3$
(pure factorization). The results for the CP-conjugate process are
given in parentheses.}
\label{Bdphi-K}
\end{table}

\section{Summary}

In summary, we have examined T violation in $B$ decays to two vector
mesons. We find that T-violating effects in the SM are tiny, except
for a few cases with radially-excited vector mesons in the final
state. On the other hand T violation with physics beyond the SM can
be large and measurable. T-violating asymmetries are therefore
excellent probes of the presence and nature of new physics.


\begin{thebibliography}{9}

\bibitem{CPreview} For a review, see, for example, {\it The BaBar
  Physics Book}, eds.\ P.F. Harrison and H.R. Quinn, SLAC Report 504,
  October 1998.

\bibitem{Valencia} G. Valencia, \prd{39}{1989}{3339}.

\bibitem{KP} G. Kramer and W.F. Palmer, \prd{45}{1992}{193},
  \plb{279}{1992}{181}, \prd{46}{1992}{2969}; G. Kramer, W.F. Palmer
  and T. Mannel, \zpc{55}{1992}{497}; G. Kramer, W.F. Palmer and
  H. Simma, \npb{428}{1994}{77}; A.N. Kamal and C.W. Luo,
  \plb{388}{1996}{633}.

\bibitem{DDLR} A.S. Dighe, I. Dunietz, H.J. Lipkin and J.L. Rosner,
  \plb{369}{1996}{144}; B. Tseng and C.-W. Chiang, hep-ph/9905338.

\bibitem{CW} N. Sinha and R. Sinha, \prl{80}{1998}{3706}; C.-W. Chiang
  and L. Wolfenstein, \newprd{61}{2000}{074031}.

\bibitem{Chiang} The full time-dependent $B\to V_1 V_2$ angular
  distribution is discussed in C.-W. Chiang,
  \newprd{62}{2000}{014017}.

\bibitem{DL} A.~Datta and D.~London, arXiv:hep-ph/0303159.

\bibitem{BuraseffH} See, for example, G. Buchalla, A.J. Buras and
  M.E. Lautenbacher, Rev.\ Mod.\ Phys.\ {\bf 68}, 1125 (1996),
  A.J. Buras, ``Weak Hamiltonian, CP Violation and Rare Decays,'' in
  {\it Probing the Standard Model of Particle Interactions}, ed.\
  F. David and R. Gupta (Elsevier Science B.V., 1998), pp.\ 281-539.

\bibitem{datta-he} T.~E.~Browder, A.~Datta, X.~G.~He and S.~Pakvasa,
Phys.\ Rev.\ D {\bf 57}, 6829 (1998) [arXiv:hep-ph/9705320].

\bibitem{Kayser} B. Kayser, Nucl.\ Phys.\ Proc.\ Suppl.\ {\bf 13}, 487
(1990).

\bibitem{SMlambdab} W. Bensalem, A. Datta and D. London, Phys.\ Lett.\
B {\bf 538}, 309 (2002).

\bibitem{dattalip2} A. Datta, H.J. Lipkin and P.J. O'Donnell, Phys.\
Lett.\ B {\bf 529}, 93 (2002).

\bibitem{NPlambdab} W. Bensalem, A. Datta and D. London, Phys.\ Rev.\
D {\bf 66}, 094004 (2002).

\bibitem{Jpsiks} $\bd(t) \to J/\psi\ks$: B. Aubert {\it et al.}
(BABAR Collaboration), hep-ex/0207042; \\ K. Abe {\it et al.} (Belle
Collaboration), hep-ex/0208025. The weighted average of these two
results is taken from the CKMfitter web site:
http://www.slac.stanford.edu/$\sim$laplace/ckmfitter.html

\bibitem{phiks} $\bd(t) \to \phi\ks$: B. Aubert {\it et al.}  (BABAR
Collaboration), hep-ex/0207070 ($\sin 2\beta ~~[\phi \ks]_{BaBar} =
-0.19 ^{+0.52}_{-0.50} \pm 0.09$); T. Augshev (Belle Collaboration),
talk given at ICHEP 2002, BELLE-CONF-0232 ($\sin 2\beta ~~[\phi
\ks]_{Belle} = -0.73 \pm 0.64 \pm 0.18$).

\bibitem{Gross} Y. Grossman and M.P. Worah, Phys.\ Lett.\ B {\bf 395},
241 (1997).

\bibitem{LonSoni} D. London and A. Soni, Phys.\ Lett.\ B {\bf 407}, 61
(1997).

\bibitem{phiKsNP} G. Hiller, Phys.\ Rev.\ D {\bf 66}, 071502 (2002);
M. Ciuchini and L. Silvestrini, Phys.\ Rev.\ Lett.\ {\bf 89}, 231802
(2002); M. Raidal, Phys.\ Rev.\ Lett.\ {\bf 89}, 231803 (2002);
J.P. Lee and K.Y. Lee, arXiv:hep-ph/0209290; S. Khalil and E. Kou,
arXiv:hep-ph/0212023; G.L. Kane, P. Ko, H.-b. Wang, C. Kolda,
J.H. Park and L.T. Wang, arXiv:hep-ph/0212092; S.-w. Baek,
arXiv:hep-ph/0301269; C.W. Chiang and J.L. Rosner,
arXiv:hep-ph/0302094.

\bibitem{dattarparity} A. Datta, Phys.\ Rev.\ D {\bf 66}, 071702
(2002).

\bibitem{bhaskar} B. Dutta, C.S. Kim and S. Oh, Phys.\ Rev.\ Lett.\
{\bf 90}, 011801 (2003); A. Kundu and T. Mitra, arXiv:hep-ph/0302123.

\bibitem{dattalip3} See, for example, A. Datta, H.J. Lipkin and
P.J. O'Donnell, Phys.\ Lett.\ B {\bf 540}, 97 (2002), Phys.\ Lett.\ B
{\bf 544}, 145 (2002).

\bibitem{LSS2} In the same spirit, for a comparison of $\bd \to J/\psi
  \ks$ and $B \to J/\psi K^*$, see D. London, N. Sinha and R. Sinha,
  arXiv:hep-ph/0207007.

\bibitem{PDG} D.E. Groom et al.\ (Particle Data Group), Eur.\ Phys.\
J.\ {\bf C15} (2000) 1.

\end{thebibliography}
\end{document}